\documentclass[prb,preprint]{revtex4}
\usepackage{amsmath}

\begin{document}

\title{Interpretation of the cosmological metric}

\author{Richard J.\ Cook}
\email{richard.cook@usafa.edu}
\affiliation{Department of Physics, U.S. Air Force 
Academy, Colorado Springs, Colorado 80908}

\author{M.\ Shane Burns}
\email{sburns@coloradocollege.edu}
\affiliation{Department of Physics, Colorado College, Colorado Springs, Colorado 80903}


\begin{abstract}
The cosmological Robertson-Walker metric of general relativity is often said to have the consequences that (1) the recessional velocity $v$ of a galaxy at proper distance $\ell$ obeys the Hubble law $v=H\ell$, and therefore galaxies at sufficiently great distance $\ell$ are receding faster than the speed of light $c$; (2) faster than light recession does not violate special relativity theory because the latter is not applicable to the cosmological problem, and because ``space itself is receding'' faster than $c$ at great distance, and it is velocity relative to local space that is limited by $c$, not the velocity of distant objects relative to nearby ones; (3) we can see galaxies receding faster than the speed of light; and (4) the cosmological redshift is not a Doppler shift, but is due to a stretching of photon wavelength during propagation in an expanding universe. We present a particular Robertson-Walker metric (an empty universe metric) for which a coordinate transformation shows that none of these interpretation necessarily holds. The resulting paradoxes of interpretation lead to a deeper understanding of the meaning of the cosmological metric.
\end{abstract}


\maketitle

\section{Introduction\label{sec:intro}}

Several authors have questioned the standard expansion of space interpretation of the Robertson-Walker cosmological metric. Some of this discussion was stimulated by a \textit{Scientific American} article by Charles Lineweaver and Tamara Davis.\cite{Lineweaver1} In this article and other more technical publications,\cite{Davis1, Davis2} the authors discuss in a beautifully clear manner a number of common misconceptions about the physics of the expanding universe, taking as the ``right'' interpretation the generally accepted expansion of space interpretation based on the Robertson-Walker line element. We heartily recommend this work to the interested reader. Other authors such as Gr\o n and Elgar\o y,\cite{Gron} Chodorowski,\cite{Chodorowski1, Chodorowski2} Peacock,\cite{Peacock} and Morgan\cite{Morgan} have discussed alternatives to this interpretation or have argued that the standard expansion of space interpretations are incorrect.\cite{Chodorowski1} In this paper we point out that a number of these arguments are nothing more than interpretations of the same cosmological physics in a different coordinate systems (different reference frames).

One of the fundamental principles of general relativity (the principle of general covariance) states that all spacetime coordinate systems are equally valid for the description of nature, and that metrics that are related by a coordinate transformation are physically equivalent. This principle sounds simple enough, but one repeatedly finds in the literature arguments that amount to advocacy for the interpretation based on one set of coordinates to the exclusion of the interpretation that is natural when using another set of coordinates for the same set of events. It has been said that each generation of physicists must learn anew (usually the hard way) the meaning of Einstein's postulate of general covariance.

This paper considers a particular example of the Robertson-Walker metric for which a coordinate transformation shows that this metric describes a flat spacetime admitting Minkowski coordinates. This particular metric allows one to apply all of the arguments leading to the standard expanding-space interpretation. At the same time we can interpret the physics from a flat-spacetime, Minkowski perspective -- a perspective that is sometimes used in arguments intended to refute the expansion of space interpretation. The vastly different interpretations in these two coordinate systems provides a graphic example of how very different interpretations can exist in different coordinate frames and yet be physically equivalent. The seemingly irreconcilable interpretations present apparent paradoxes of a type not encountered in special relativity, and the understanding of the resolution of these paradoxes leads to a deeper understanding of the meaning of the postulate of general covariance.

John Wheeler has written ``Sad the week without a paradox, a difficulty, an apparent contradiction! For how can one then make progress?''\cite{Wheeler} One of the best ways to acquire a deep understanding of relativistic concepts is to work through and resolve the paradoxes of relativity (problems that, when approached in two different ways, give apparently contradictory results). The twin paradox and the problem in which a moving Lorentz-contracted pole fits into a barn in the barn's reference frame, but not in the pole's frame are problems of this type.\cite{Hartle,Taylor} The latter problem brings home the meaning of the relativity of simultaneity better than any abstract discussion of how clocks are synchronized. We hope that the following discussion of the paradoxes in cosmological theory will lead to such a learning experience.

We first give a more quantitative discussion of some of the ``correct'' conceptual pictures of cosmological phenomena addressed by Lineweaver and Davis,\cite{Lineweaver1} views that are held to be correct by most cosmologists. This picture may be viewed as the accepted conventional interpretation derived from the Robertson-Walker metric. We then discuss a particular model universe (an empty Robertson-Walker universe) which, when transformed to a manifestly flat-spacetime form, seems to contradict the conventional interpretation. Finally, we discuss the essential insights required to resolve the various contradictions and thereby obtain a deeper understanding of the Robertson-Walker spacetime in this and other contexts.

\section{The Robertson-Walker Metric\label{sec:RWmetric}}

A standard form of the Robertson-Walker metric is
\begin{equation}
ds^{2}=-c^{2}dt^{2}+a^{2}(t)\left[\frac{dr^{2}}{1-kr^{2}}+r^{2}(d\theta^{2}+\sin^{2}\theta\, d\phi^{2})\right],
\label{RWmetricNo1}
\end{equation}
where the spatial geometry has positive, zero, or negative curvature for $k=+1$, $0$, and $-1$, respectively, and $a(t)$ is the scale factor describing the expansion of the universe. Galaxies reside at constant values of the comoving coordinates $r$, $\theta$, $\phi$, and the time coordinate $t$ is the proper time on clocks moving with the galaxies.

For our purpose a different radial coordinate,
\begin{equation}
\chi=\!\int_{0}^{r}\frac{dx}{\sqrt{1-kx^{2}}},
\label{NewRadial}
\end{equation}
is convenient. For this radial marker the metric in Eq.~(\ref{RWmetricNo1}) becomes
\begin{equation}
ds^{2}=-c^{2}dt^{2}+a^{2}(t)\left[d\chi^{2}+\Sigma_k^{2}(d\theta^{2}+\sin^{2}\theta\, d\phi^{2})\right],
\label{RWmetricNo2}
\end{equation}
where $\Sigma_k=\sin\chi$ for $k=+1$, $\Sigma_k=\chi$ for $k=0$, and $\Sigma_k=\sinh\chi$ for $k=-1$.

The Robertson-Walker metric is the most general metric describing a spatially homogeneous and isotropic spacetime. The form of this metric is derived from these symmetries alone, and not from the Einstein field equation. If the Robertson-Walker metric is to be a solution of Einstein's equation, then $a(t)$ must satisfy the Friedmann equation; the equation describing the gravitational dynamics of a spatially homogeneous spacetime.

\section{The Conventional Interpretation\label{sec:Standard}}

The following are the generally accepted points of interpretation of the Robertson-Walker metric discussed by Lineweaver and Davis,\cite{Lineweaver1} but presented here in a more quantitative form.

\subsection{Distant Objects Recede Faster Than Light} 
From the Robertson-Walker metric in Eq.~(\ref{RWmetricNo2}) we see that the proper distance from an observer at $\chi=0$ to a galaxy at comoving coordinate $\chi$ (measured at a given time $t$) is $\ell=a(t)\chi$. It follows that the recessional velocity is given by Hubble's law
\begin{equation}
v\equiv\frac{d\ell}{dt}=H(t)\ell,
\label{HubbleLaw}
\end{equation}
where
\begin{equation}
H(t)=\frac{1}{a}\frac{da}{dt}
\label{HubbleParameter}
\end{equation}
is the Hubble parameter. Equation (\ref{HubbleLaw}) indicates that, at distances greater than the Hubble distance,
\begin{equation}
\ell_{H}=\frac{c}{H(t)},
\label{HubbleDistance}
\end{equation}
the recessional velocity exceeds $c$ in an expanding universe. 

\subsection{Space Itself is Expanding} 
The world lines of constant $\chi$, which mark the positions of distant galaxies, are geodesics of the Robertson-Walker metric (\ref{RWmetricNo2}), and the non-rotating Cartesian axes following these paths are local inertial reference frames for all time $t$. The proper distance $d\ell=a(t)d\chi$ between two such frames at $\chi=\mbox{constant}$ and $\chi+d\chi=\mbox{constant}$ increases with time in an expanding universe, and the integrated proper distance between the local inertial observer frame at $\chi=0$ and the local galaxy frame at $\chi=\mbox{constant}$, namely $\ell=a(t)\chi$, increases with time according to the Hubble law (\ref{HubbleLaw}). Thus the two local inertial frames are separating at a rate exceeding the speed of light when the proper distance between them exceeds the Hubble distance $\ell_{H}$. This increasing separation of local inertial frames is referred to as the expansion of space itself, and, except for small deviations, the matter of the universe follows this expansion. The velocity of any object in the local inertial frame at its location is less than $c$. We say that space itself is expanding because, in a non-expanding space such as the Minkowski space of special relativity, the relative velocity of material objects cannot exceed $c$ for any spatial separation of those objects. 

\subsection{We Can See Galaxies Receding Faster than Light}\label{sec:SeeGal}
The statement that we can see galaxies receding faster than light is sometimes thought to be false because the local inertial frame of such a galaxy is moving away from us faster than light, and so photons emitted by the galaxy in our direction are ``dragged away'' and make no progress in our direction. This interpretation is correct initially, but as time progresses, the Hubble distance (\ref{HubbleDistance}) increases and eventually the photons which are attempting to reach us are no longer in a local inertial frame receding faster than light. The photons then decrease their distance from us and eventually reach our position ($\chi=0$), allowing us to see a galaxy that is receding faster than light. 

The mathematics of galaxy recession proceeds as follows. The rate of change of the proper distance $\ell=a(t)\chi$ for any moving object is
\begin{equation}
\frac{d\ell}{dt}=\frac{da}{dt}\chi+a\frac{d\chi}{dt}.
\label{Dproper}
\end{equation}
The first term on the right is the Hubble-expansion term $H(t)\ell$. For galaxies with $\chi=\chi_{G}=\mbox{constant}$ the second term vanishes, and Eq.~(\ref{Dproper}) is just the Hubble expansion law $d\ell_{G}/dt=H(t)\ell_{G}$. In contrast, for photons propagating radially, the null condition $ds^{2}=-c^{2}dt^{2}+a^{2}(t)d\chi^{2}=0$ gives $a(t)d\chi/dt=\pm c$ for the second term, and Eq.~(\ref{Dproper}) for photons propagating toward us becomes
\begin{equation}
\frac{d\ell_{\gamma}}{dt}=H(t)\ell_{\gamma}-c.
\label{PhotonProp}
\end{equation}
This equation states that the photon is ``swept away'' by the Hubble expansion [the first term $H(t)\ell_{\gamma}$] while at the same time propagating toward us at speed $c$ (the second term). 

As an example, if the scale factor increases linearly with time, $a(t)=\alpha t$, where $\alpha$ is constant,\cite{footnote} then the Hubble parameter is $H(t)=1/t$, and the Hubble distance (\ref{HubbleDistance}) is $\ell_{H}=c\,t$; that is, the sphere of this radius beyond which galaxies recede faster than light moves outward at speed $c$. Consider a galaxy at distance $\ell_{G}=a(t_{e})\chi_{G}$ greater than the Hubble distance $\ell_{H}=c\,t_{e}$ at the time $t_{e}$ photons are emitted. Such a galaxy forever recedes at a velocity $v_{G}=(da/dt)\chi_{G}=\alpha\chi_{G}=\mbox{constant}$ which is greater than $c$. In contrast, photons emitted by this galaxy in our direction propagate according to Eq.~(\ref{PhotonProp}), which, for the present case, is
\begin{equation}
\frac{d\ell_{\gamma}}{dt}=\frac{\ell_{\gamma}}{t}-c.
\label{Ray}
\end{equation}
The solution for photons emitted at a distance $\ell_{G}$ at time $t_{e}$,
\begin{equation}
\ell_{\gamma}(t)=\frac{\ell_{G}t}{t_{e}}-c\,t\ln\left(\frac{t}{t_{e}}\right),
\label{Solution}
\end{equation}
shows that the photons eventually reach us at $\ell_\gamma=0$ at time $t=t_{e}\exp(\ell_{G}/c\,t_{e})$. Immediately after emission the photons are ``dragged into recession'' by the Hubble expansion, but when the Hubble sphere of radius $\ell_{H}=c\,t$ expands to include these photons [which occurs at time $t=t_{e}\exp(\ell_{G}/c\,t_{e}-1)$], the photon velocity (\ref{Ray}) switches from recession ($d\ell_{\gamma}/dt>0$) to approach ($d\ell_{\gamma}/dt<0$), allowing us to eventually see the galaxy that is forever receding faster than light.

\subsection{The Cosmic Redshift is Not a Doppler Shift} 
A standard argument that the cosmic redshift is not a Doppler shift shows that the cosmological redshift for the Robertson-Walker metric has the form
\begin{equation}
z=\frac{\lambda_{r}-\lambda_{e}}{\lambda_{e}}=\frac{a(t_{r})}{a(t_{e})}-1.
\label{Redshift}
\end{equation}
That is, it depends only on the scale factor $a(t)$ at the time $t_{e}$ of emission and the time $t_{r}$ of reception.\cite{Hartle2} Equation~\eqref{Redshift} is not the formula for the Doppler shift of a source receding with the Hubble velocity $v=H(t)\ell$. There would still be a cosmological redshift if the universe were not expanding at all at the time of emission and at the time of reception [$v(t_{e})=0$ and $v(t_{r})=0$], provided that some expansion had occurred during the intervening time so that $a(t_{r})>a(t_{e})$. For this reason the cosmological redshift is interpreted, not as a Doppler effect, but as a stretching of the wavelength of light during propagation when the universe is expanding.

We note that a special-relativistic Doppler effect at the Hubble recessional velocity $v=H(t)\ell=(da/dt)\ell/a$ would give a redshift of
\begin{equation}
z=\sqrt{\frac{c+v}{c-v}}-1=\sqrt{\frac{c+H\ell}{c-H\ell}}-1=\sqrt{\frac{ca+\dot{a}\ell}{ca-\dot{a}\ell}}-1,
\label{DopplerZ}
\end{equation}
an expression entirely different from the cosmological redshift (\ref{Redshift}), and one that does not even give a real number for $z$ at distances $\ell$ greater than the Hubble distance $\ell_{H}$ of Eq.~(\ref{HubbleDistance}).

\subsection{Special Relativity Does Not Apply in an Expanding Universe} 
The conclusion that special relativity does not apply in an expanding universe is reached from a number of different arguments. A recessional velocity greater than $c$ of material objects (galaxies), the failure of the special-relativistic Doppler formula to account for the cosmological redshift, the stretching of photon wavelength during propagation, and the size of the observable universe being greater than $c$ times the age of the universe all argue for a failure of special-relativistic ideas.

\subsection{\label{sec:k}The Curvature of Three-Dimensional Space is Determined by the Value of $k$ in the Robertson-Walker Metric}
The curvature of three-dimensional space as determined by the value of $k$ in the Robertson-Walker metric is not addressed in Ref.~\onlinecite{Lineweaver1} but is a widely held belief in the cosmological community. The curvature scalar for the cosmological three-dimensional space with the metric
\begin{equation}
d\ell^{2}=a^{2}(t)\left[d\chi^{2} +\Sigma_k^{2}(d\theta^{2}+\sin^{2}\theta d\phi^{2})\right],
\label{ThreeMetric}
\end{equation}
is\cite{Misner}
\begin{equation}
^3\mathcal{R}=\frac{6k}{a^{2}(t)}.
\label{ThreeCurvature}
\end{equation}
Hence, we may conclude that the spatial curvature $^3\mathcal{R}$ is positive if $k=+1$, negative if $k=-1$, and is zero if $k=0$.

We emphasize that all of these conclusions are derived solely from the form of the Robertson-Walker metric and have nothing specifically to do with the Einstein or Friedmann equations which determine how $a(t)$ changes with time for a universe with a specified matter and radiation content.

\section{An Alternative Interpretation}

There is one and only one instance of the Robertson-Walker metric for which measurement in an expanding universe can be directly compared with measurement in the static flat spacetime of special relativity, namely, the negative spatial curvature ($k=-1$) empty universe ($T_{\mu\nu}=0$) case. In this case the Friedmann equation yields the scale factor
\begin{equation}
a(t)=c t.
\label{ScaleFactor}
\end{equation}
That is, the scale factor is proportional to time with a proportionality constant equal to the speed of light $c$. For this scale factor the Hubble parameter (\ref{HubbleParameter}) is $H(t)=1/t$, and, at the present time $t_{0}$, the observed value of $H_{0}=H(t_{0})\approx 72$\,(km/s)/Mpc places the age of this model universe at the Hubble time $t_{0}=t_{H}=1/H_{0}$ of about $14$ billion years. This model universe does not describe the universe in which we live. Its purpose is to serve as a counterexample to the conventional interpretations of the Robertson-Walker metric. Our model universe has the Robertson-Walker metric
\begin{equation}
ds^{2}=-c^{2}dt^{2}+c^{2}t^{2}\left[d\chi^{2}+\sinh^{2}\chi(d\theta^{2} +\sin^{2}\theta d\phi^{2})\right].
\label{ModelMetric}
\end{equation}
We emphasize that all of the previous arguments leading to the conventional points of interpretation of the Robertson-Walker metric apply to this metric. However, this metric is special in that it is a flat-spacetime metric. The coordinate transformation that brings Eq.~(\ref{ModelMetric}) into a manifestly flat form is
\begin{align}
c t&=\sqrt{(c\,\bar{t})^{2}-\bar{r}^{2}}, \label{Ttrans} \\
\chi&=\tanh^{-1}\left(\frac{\bar{r}}{c\,\bar{t}}\right), \label{Rtrans}
\end{align}
or its inverse
\begin{align}
\bar{t}&=t\cosh\chi, \label{Ttransinverse} \\
\bar{r}&=c t\sinh\chi. \label{Rtransinverse} 
\end{align}
The metric in Eq.~(\ref{ModelMetric}) becomes
\begin{equation}
ds^{2}=-c^{2}d\bar{t}^{2}+d\bar{r}^{2}+\bar{r}^{2}(d\theta^{2}+\sin^{2}\theta d\phi^{2}),
\label{FlatMetric}
\end{equation}
where $\bar{r}$ is the proper distance measured from the origin $\bar{r}=0$ in the new reference frame, and $\bar{t}$ is the proper time on synchronized clocks at rest ($\bar{r}, \theta, \phi =\mbox{constant}$) in these spherical-polar coordinates. This metric describes a static, flat spacetime in which all measurements have their well-understood meanings familiar from special relativity. The recessional motion of galaxies in this flat space is essentially the same as that in the Milne cosmological model.\cite{Milne2}

We shall refer to the expanding reference frame with local observers at fixed values of the comoving coordinates $\chi$, $\theta$, $\phi$, with clocks at their locations to measure time $t$ as the ``expanding frame'' $K$; we shall refer to the frame with local observers at fixed values of the coordinates $\bar{r}$, $\theta$, $\phi$, and with clocks at their locations to measure time $\bar{t}$ as the ``rigid frame'' $\bar{K}$.

We next evaluate each of the previous conventional interpretations concerning the Robertson-Walker metric (\ref{ModelMetric}) discussed in Sec.~III from the point of view of the rigid frame $\bar{K}$.

\subsection{Do Distant Objects Recede Faster Than Light?}
For objects at rest in the expanding frame $K$ ($\chi$, $\theta$, and $\phi$ constant) the proper distance $\ell=a(t)\chi=c t\chi$ increases according to the Hubble law $d\ell/dt=H(t)\ell=\ell/t$, just as for any other Robertson-Walker metric, and this velocity exceeds $c$ for $\ell$ greater than the Hubble distance $\ell_{H}=c t$. In the rigid frame $\bar{K}$ the recessional velocity calculated from Eqs.~(\ref{Ttransinverse}) and (\ref{Rtransinverse}) for $\chi=\mbox{constant}$, namely,
\begin{equation}
\bar{v}=\frac{d\bar{r}}{d\bar{t}}=c\tanh\chi,
\label{ReVelocityBar}
\end{equation}
never exceeds $c$, and does not obey the Hubble law of recession. We conclude that the Hubble law is a frame-dependent, or coordinate-dependent effect, and is not a coordinate-independent law.

These differing values for the recessional velocity come about as follows. Consider the sequence of local inertial reference frames in the expanding frame $K$, each at a particular $\chi=\mbox{constant}$ and separated by equal coordinate displacements $d\chi$ extending from $\chi=0$ to some distant galaxy at $\chi_{G}=\mbox{constant}$. The observers of frame $K$ are at rest in these local inertial frames and measure distance $d\ell=a(t)d\chi$ and relative velocity $dv=\dot{a}(t)d\chi$ between adjacent frames. By adding all these distances and velocities together, the expanding-frame observers conclude that the proper distance to $\chi_{G}$ is $\ell=\!\int\! d\ell=a(t)\!\int\! d\chi=a(t)\chi_{G}$, and its recessional velocity is $v=\!\int\! dv=\dot{a}(t)\!\int\! d\chi=\dot{a}(t)\chi_{G}$. These relations combine to give Hubble's law, $v=H(t)\ell$ with $H(t)=\dot{a}(t)/a(t)$, even in the present special-relativistic context.

The observers of rigid frame $\bar{K}$ object to this logic. They agree that the comoving observer of frame $K$ in one local inertial frame is moving away from the comoving observer in the next local inertial frame, but they argue that these two frames are connected by a Lorentz transformation, and therefore, to go from the local comoving frame at $\chi=0$ to the one at $\chi_{G}$, one must successively apply each of these Lorentz transformations to determine the velocity of the frame at $\chi_{G}$ relative to the one at $\chi=0$. The resulting relative velocity cannot exceed $c$ because no combinations of Lorentz transformations can yield a velocity of the final frame relative to the first greater than $c$.

The rigid-frame observers point out to the expanding-frame observers that if they ``correctly'' apply infinitesimal Lorentz transformations between neighboring local inertial frames for each coordinate displacement $d\chi$ from $\chi=0$ to $\chi=\chi_{G}$ at a given time $t$, the velocity of the composite Lorentz transformation, namely the velocity of the galaxy at $\chi_{G}$ relative to the observer at $\chi=0$, is $\bar{v}=c\tanh[a(t)\!\int_{0}^{\chi_{G}}d\chi]=c\tanh[v/c]$, where $v$ is the expanding-frame recessional velocity. This result, which is the velocity of Eq.~(\ref{ReVelocityBar}) for the case [$a(t)=c t$, $\ell=c t\chi$] under consideration, never exceeds $c$ as it cannot in a flat spacetime. The rigid-frame observers might also attempt to convince the expanding-frame observers that by adding together the relative velocities $dv=\dot{a}(t)d\chi$ of adjacent local inertial frames to obtain the velocity of a far off galaxy, they are actually using the classical addition law for velocities instead of the correct relativistic velocity-addition law. It is this error that leads them to conclude that galaxies can recede faster than light. 

\subsection{Is Space Expanding?}
No one would seriously suggest that the rigid frame $\bar{K}$ is expanding. From the point of view of frame $\bar{K}$ the coordinate grid of frame $K$ is expanding, and the $K$-frame observers, which are fixed to the $K$-frame grid, are separating from one another, but the phrase ``space itself is expanding'' seems inappropriate from the $\bar{K}$ observers' point of view.

\subsection{Can We See Objects Receding Faster Than Light?}
In rigid frame $\bar{K}$, no material object, including galaxies, travels faster than light. In this frame the recessional velocity (\ref{ReVelocityBar}) never exceeds $c$, and so we can see galaxies that were ``incorrectly'' assigned superluminal velocities by the expanding-frame observers. But the set of objects truly traveling faster than $c$ is empty, and the question is meaningless.

\subsection{Can the Cosmological Redshift be a Doppler Shift?}
For our model expanding universe, $a(t)=c t$, the cosmological redshift, Eq.~(\ref{Redshift}), reads
\begin{equation}
z=\frac{t_{r}}{t_{e}}-1.
\label{ModelRedshift}
\end{equation}
For radial light propagation the Robertson-Walker metric (\ref{ModelMetric}) gives the null condition $ds^{2}=-c^{2}dt^{2}+c^{2}t^{2}d\chi^{2}=0$, which integrates to $t_{r}/t_{e}=\exp(\chi_G)$, where $\chi_G$ is the comoving coordinate of the source galaxy that emitted the light at time $t_e$. Hence, the redshift for this galaxy is
\begin{equation}
z=\exp(\chi_G)-1,
\label{ExpandRedshift}
\end{equation}
which is the redshift in terms of the expanding-space comoving coordinates.

From the rigid-frame, special-relativistic perspective, space is static and the point $\chi=\mbox{constant}$ recedes with velocity $\bar{v}=c\tanh\chi$. This recessional velocity produces a special-relativistic Doppler shift
\begin{equation}
z=\sqrt{\frac{c+\bar{v}}{c-\bar{v}}}-1=\sqrt{\frac{1+\tanh\chi}{1-\tanh\chi}}-1=\exp(\chi)-1.
\label{SRDoppler}
\end{equation}
Thus, for this case the Robertson-Walker redshift for a source receding faster than light in frame $K$ is, according to the observers in frame $\bar{K}$, actually a special-relativistic Doppler effect with no source exceeding the speed of light.

\subsection{Does Special Relativity Apply in an Expanding Universe?}
The Robertson-Walker metric (\ref{ModelMetric}) with scale factor $a(t)=c t$ would certainly qualify as an expanding universe because the proper distance $\ell=a(t)\chi$ to comoving objects ($\chi=\mbox{constant}$) is increasing. But, as we have seen, a coordinate transformation shows this metric to be one possible representation of the flat spacetime of special relativity. Therefore special relativity certainly applies to this particular Robertson-Walker universe, in contradiction to the conclusion reached in Sec.~\ref{sec:k}.

Note also that in the rigid frame $\bar{K}$, the redshift is entirely a Doppler shift that can be measured by any observer at rest ($\bar{r}=\mbox{constant}$) and closer to $\bar{r}=0$ than the source; that is, there is no stretching of the wavelength as photons propagate in this rigid frame. We conclude that wavelength stretching during light propagation in an expanding universe is a coordinate-dependent effect, and not an invariant property.

\subsection{The Curvature of Three-Dimensional Space}
The model Robertson-Walker universe in Eq.~(\ref{ModelMetric}) has $k=-1$ and negative spatial curvature [spatial curvature scalar $^3\mathcal{R}=-6/a^{2}(t)$] in the expanding reference frame $K$. And yet the coordinate transformation Eqs.~(\ref{Ttrans}) and (\ref{Rtrans}) shows this spacetime to be flat with a flat spatial metric
\begin{equation}
d\bar{\ell}^{2}=d\bar{r}^{2}+\bar{r}^{2}(d\theta^{2}+\sin^{2}\theta d\phi^{2}) 
\label{FlatSpace}
\end{equation}
in the rigid reference frame $\bar{K}$. We conclude that the curvature of three-dimensional space depends on the choice of spacetime coordinates, and is not an invariant property determined by the value of $k$ in the expanding frame.

\section{Resolution of Paradoxes}

The special relativistic metric (\ref{FlatMetric}) describes the same spacetime as the Robertson-Walker metric (\ref{ModelMetric}). The analysis of Sec.~IV seems to leave us with several paradoxes. Our quandary is summarized in Table~\ref{tbl:paradox}. 

What can be learned from these paradoxical results? First off, we reemphasize that this model universe is not intended to describe the actual universe in which we live. This model describes an empty universe, as does any flat spacetime metric. The example Robertson-Walker metric in Eq.~(\ref{ModelMetric}) is expected to correctly describe a universe in which the density of matter and/or radiation is so small that it has a negligible effect on the flat spacetime geometry (this assumption is routinely made in special relativity when particle motion and the propagation of radiation is studied while neglecting their effect on the metric). The galaxies in our model universe are comoving in the expanding frame $K$; they need only be of exceedingly small mass. In this context the flat-space Robertson-Walker metric in Eq.~(\ref{ModelMetric}) has a conventional interpretation that is strikingly at variance with the special relativity metric in Eq.~~(\ref{FlatMetric}).

These paradoxes cannot be avoided by considering a more realistic universe containing matter and radiation. For any Robertson-Walker universe (curved or flat), we can choose a new rigid radial coordinate $\bar{r}$ for which fiducial observers residing at constant values of this coordinate have an unchanging radial distance between them, and all of the paradoxes occur in these cases also, except that the redshift has, perhaps, a component attributed to gravitational time dilation as well as the Doppler effect, depending on one's interpretation. The paradoxes result from the different spacetime coordinates used in the Robertson-Walker and rigid-coordinate cases, and not from the presence or absence of matter and radiation. The flat-space case was chosen for simplicity and because the essential physics is presented most clearly in this simple case for which the argument is not obscured by inessential curvature effects or cumbersome mathematics.

The all important lesson to be learned is not that one or the other of these reference frames gives the right description of cosmological events, but that both are entirely correct descriptions of cosmological events for their respective sets of fiducial observers.
For the observers fixed to the comoving coordinates of a Robertson-Walker reference frame, even in flat space, the conventional interpretation of the associated Robertson-Walker metric (with galaxies receding faster than light, and so on) is appropriate. And for fiducial observers on a rigid cartesian frame in flat space, the principles of special relativity (such as $c$ being a limiting velocity) apply. There is no contradiction or inconsistency, only different descriptions of the same physics for different sets of observers (different reference frames). 

Many of the paradoxes can be resolved by relating the length and time intervals of the observers in the two reference frames. Consider two observers, one in the expanding frame $K$ and the other in the rigid frame $\bar{K}$, who happen to be at the same place at a given time. Both of these observers use meter sticks and standard clocks at rest in their respective local comoving frames to measure proper lengths and proper times. But these two frames are in relative motion, and so the distances and times of these observers are related by a Lorentz transformation at the relative velocity of the two observers. Therefore, we should, at a minimum, expect to encounter all of the seeming paradoxes associated with the Lorentz transformation of special relativity. In fact, the paradoxes we encounter when comparing $K$ and $\bar{K}$ observations are even more difficult to unravel since the relative velocity of observers in the two frames depends on place and time, and observers stationed at fixed comoving coordinates of the expanding frame are in relative motion. 

From these considerations it is evident that proper distance (the sum of locally measured proper distances) should not be the same in the two frames because proper length is not Lorentz invariant even in special relativity because it experiences Lorentz contraction. Similarly, time is not Lorentz invariant because it undergoes motional time dilation, and simultaneity is not invariant either. There is no reason to believe that the size of the observable universe (a proper distance) or recessional velocity (a proper distance divided by a proper time) should be the same in different reference frames. The surprise, in our cosmological example is just how vastly different these measurements can be in the two different reference frames.

To make these ideas quantitative note that the fiducial observers of the expanding frame $K$ [metric (\ref{ModelMetric})] each have local comoving inertial frames with short proper distances measured from a given observer in the $\chi$, $\theta$, and $\phi$ directions given by
\begin{subequations}
\begin{align}
d\ell^{\chi}&=c\,t\,d\chi, \\
d\ell^{\theta}&=c\,t\sinh\chi\, d\theta, \\
d\ell^{\phi}&=c\,t\sinh\chi\sin\theta\, d\phi, \label{PropersK}
\end{align}
\end{subequations}
and a local time coordinate $dx^{0}=c\,dt$. These are the local Minkowski coordinates for the comoving observer with coordinates $\chi$, $\theta$, $\phi = \mbox{constant}$ [the metric in Eq.~(\ref{ModelMetric}) in terms of these local coordinates is $ds^{2}=-(dx^{0})^{2}+(d\ell^{\chi})^{2}+(d\ell^{\theta})^{2}+(d\ell^{\phi})^{2}$]. Similarly, the local inertial frames of the fiducial observers of the rigid frame $\bar{K}$ [the metric in Eq.~(\ref{FlatMetric})] have space coordinates
\begin{subequations}
\begin{align}
d\bar{\ell}^{\bar{r}}&=d\bar{r}, \\
d\bar{\ell}^{\bar{\theta}}&=\bar{r}\,d\theta, \\
d\bar{\ell}^{\phi}&=\bar{r}\sin\theta\, d\phi.\label{PropersKBAR}
\end{align}
\end{subequations}
The time coordinate $d\bar{x}^{0}=c d\bar{t}$ in terms of these local coordinates is $ds^{2}=-(d\bar{x}^{0})^{2}+(d\bar{\ell}^{\bar{r}})^{2}+(d\bar{\ell}^{\theta})^{2}+(d\bar{\ell}^{\phi})^{2}$]. The two local inertial frames for observers at the same place at a given time are in relative motion with the relative velocity $\bar{v}=c\tanh\chi$ [Eq.~(\ref{ReVelocityBar})] in the radial direction, and therefore the coordinates of the two local inertial frames are related by the Lorentz transformation
\begin{subequations}
\begin{align}
d\bar{x}^{0}&=\frac{dx^{0}-(\bar{v}/c)d\ell^{\chi}}{\sqrt{1-\bar{v}^{2}/c^{2}}}, \label{Trans1} \\
d\bar{\ell}^{\bar{r}}&=\frac{d\ell^{\chi}-(\bar{v}/c)dx^{0}}{\sqrt{1-\bar{v}^{2}/c^{2}}}, \label{Trans2} \\
d\bar{\ell}^{\theta}&=d\ell^{\theta}, \label{Trans3} \\
d\bar{\ell}^{\phi}&=d\ell^{\phi}. \label{Trans4}
\end{align}
\end{subequations}
By using this transformation we can relate the physical predictions in the two frames. For example, the distance to a galaxy at $\chi=\mbox{constant}$ in the expanding frame $K$ is the sum of local proper distances $d\ell^{\chi}$ at constant time $t$ ($dx^{0}=cdt=0$), namely $\ell^{\chi}=ct\chi$, and the proper distance to the same galaxy in frame $\bar{K}$ is the integral of $d\bar{\ell}^{\bar{r}}$ at constant time $\bar{x}^{0}$, namely $\bar{r}=c\bar{t}\tanh\chi$, which is more easily read directly from the coordinate transformation Eqs.~(\ref{Ttransinverse}) and (\ref{Rtransinverse}). Each of the observers of either frame, observing the momentarily co-located observer of the other frame, finds that observer's lengths Lorentz contracted in the radial direction ($\chi$ or $\bar{r}$ direction) and no contraction in the transverse directions (the $\theta$ and $\phi$ directions), and each measures the clock of the other slowed by motional time dilation.

These observations allow us to understand how the curvature of three-dimensional space can be different in the two reference frames. Consider a circle $\bar{r}=\mbox{constant}$ drawn in the $\theta=\pi/2$ plane at time $\bar{t}$ in the rigid frame $\bar{K}$. In this frame three-dimensional space is flat Euclidean space and all of the theorems of Euclidean geometry hold. In particular, the circumference $\bar{C}$ of the circle is found to be $\pi$ times the diameter $\bar{D}$ ($\bar{C}=\pi \bar{D}$). To see how these results differ from those of the expanding frame, picture the measuring rods of the rigid frame placed end-to-end along the diameter and circumference of the circle $\bar{r}=\mbox{constant}$. These rods are at rest in the rigid frame, but are moving radially relative to the fiducial observers of the expanding frame (they are moving toward $\chi=0$ on the coordinates $\chi$, $\theta$, $\phi$). Therefore, the fiducial observers of the expanding frame measure rods along the diameter to be Lorentz contracted (because each of these rods is moving in the direction of its length), but measure no contraction of the rods along the circumference (because these rods move perpendicular to their lengths). The expanding-frame observers conclude that the rigid-frame measurement of the circumference gave the correct value ($C=\bar{C}$), but the rigid frame measurement of the diameter was in error because contracted rods were used. If ``full-length'' rods had been used (rods moving with the observers of the expanding frame), then the ``correct'' diameter of this circle would be less than that measured in the rigid frame ($D<\bar{D}$), because fewer of the full-length rods fit into that interval. It follows that if $\bar{C}=\pi \bar{D}$ is the relation between diameter and circumference in the rigid frame, and $C=\bar{C}$ and $D<\bar{D}$, then in the expanding frame we have $C>\pi D$. A space in which the circumference of a circle exceeds $\pi$ times the diameter is a space of negative curvature. The actual relation between circumference $C$ and diameter $D$ in the expanding frame is most easily calculated from the spatial metric
\begin{equation}
d\ell^{2}=c^{2}t^{2}\left[d\chi^{2}+\sinh^{2}\chi(d\theta^{2} +\sin^{2}\theta d\phi^{2})\right],
\label{SpatialMetric}
\end{equation}
and the result is
\begin{equation}
C=\left(\frac{\sinh\chi}{\chi}\right) \pi D,
\label{CDrelation}
\end{equation}
where $\chi$ is the radial coordinate of the circle in this frame.

This argument is reminiscent of Einstein's early reasoning showing that the spatial geometry on a rotating disk is non-Euclidean.\cite{Einstein1} In this case measuring rods on the circumference are Lorentz contracted due to their rotation relative to inertial space, whereas rods along the diameter are not contracted. Hence, $C>\pi D$ for measurements made by observers on the rotating disk. Again we have negative curvature for this three-space, whereas the relation $\bar{C}=\pi \bar{D}$ is obtained in the non-rotating inertial space.

Consider next the question of recessional velocity. It is just as correct in general relativity to say that distant galaxies can recede with superluminal velocities as it is to say they cannot recede faster than $c$ -- it depends on which reference frame is being used. Recessional velocity is a coordinate-dependent concept and is not the same for different observers. The velocity of an object depends on the reference frame in which it is observed even in classical mechanics and vanishes in the comoving frame. For our cosmological example the essential difference between special and general relativity is that in the latter we often use reference frames in which the fiducial observers are in relative motion, whereas in special relativity we are accustomed to using reference frames for which the distances between fiducial observers remain rigidly fixed. The difference in recessional velocities for the cosmological frames considered here may be attributed to this difference. In both frames $c$ is a limiting velocity for a local observer. In a rigid reference frame $c$ is a limiting velocity even at a distance from the observer, as in special relativity. But, in an expanding reference frame the distance between neighboring fiducial observers is increasing, and the sum of such relative motions over large distances increases without bound. The velocity of a distant object, though limited locally, can have any value for a sufficiently distant observer because the local velocity and the relative velocity of local and distant fiducial observers add to give the velocity of the object, for example, a galaxy relative to the distant observer. In special relativity $c$ is a local and a global limiting velocity because the relative velocity of fiducial observers in a Minkowski frame is zero.

Enough has been said to resolve any remaining paradoxes related to rigid and expanding reference frames. There are no true paradoxes because general relativity is a consistent theory; there are only different interpretations provided by different reference frames. The example given here illustrates how vastly different such physical pictures can be. This difference is a necessary consequence of the principle of general covariance. The principle of general covariance (that we can use any spacetime coordinates for the description of nature) applies just as well to the flat spacetime of special relativity as it does to curved spacetimes, and once this concept is understood, the physical equivalence of the expanding and rigid frames for our flat model universe becomes evident. The only danger is believing that the interpretations in a particular reference frame, such as the conventional interpretations of the Robertson-Walker metric so well presented by Lineweaver and Davis,\cite{Lineweaver1} are invariant properties, which they are not. Most, if not all of the interesting phenomena described in Ref.~\onlinecite{Lineweaver1} are specific to the conventionally chosen comoving Robertson-Walker reference frame and the fiducial observers affixed to this frame. Observations by a different set of observers (a different reference frame or different spacetime coordinates) can lead to very different (but equally correct and physically equivalent) descriptions of cosmic phenomena.

The reader who is interested in pursuing further these or other topics in cosmology is referred to the excellent texts by Weinberg,\cite{Weinberg1, Weinberg2} Peebles,\cite{Peebles} and Carroll \cite{Carroll} in addition to the work we have already cited.

\begin{acknowledgments}
The authors wish to thank Antonio Mondragon for his valuable comments and suggestions. We also wish to thank \O yvind Gr\o n for pointing out an error in an early version of the manuscript and thanks also to M.\ J.\ Chodorowski for useful comments contrasting his and our views on the cosmological interpretation.
\end{acknowledgments}

\section*{Table}

\centering
\begin{table}[htbp]
\begin{tabular}{lccc} 
\hline
\hline
Interpretation & Expanding Frame ($K$) & & Rigid Frame ($\bar{K}$) \\
\hline
Distant galaxies recede faster than light	& Yes & & No \\
The Hubble law $v=H\ell$ applies &Yes & &No\\
We can see galaxies receding faster than light & Yes & & No \\
Space itself is expanding & Yes & & No \\
Cause of redshift & wavelength stretching & & Doppler effect \\
Spatial curvature & negative & & flat \\
\hline
\hline
\end{tabular}
\caption{Differences in interpretation derived from the analysis of the expanding universe frame (Robertson-Walker metric) and rigid frame (special relativity metric) for the same empty-universe spacetime. }
\label{tbl:paradox}
\end{table}


\begin{thebibliography}{99}
\bibitem{footnote}A scale factor of this form is a solution to the Friedmann equation for at least two special cases. One is a negatively curved empty universe ($k=-1$ and $T_{\mu\nu}=0$). The other is a $k=0$ universe with the equation of state $P = -\tfrac{1}{3} \rho c^2$.
\end{thebibliography}
\end{document}